\documentstyle[12pt,aaspp]{article}
\tighten
\eqsecnum
\begin{document}

\title{The X-ray Expansion of the Supernova Remnant Cassiopeia A} 
\author{Barron Koralesky\altaffilmark{1}, L. Rudnick\altaffilmark{2}} 
\affil{Department of Astronomy, University of Minnesota, 116 Church 
Street SE, Minneapolis, MN 55455} 
\author{E. V. Gotthelf\altaffilmark{3}, \& J. W. Keohane\altaffilmark{4}} 
\affil{Laboratory for High Energy 
Astrophysics, Goddard Space Flight Center, Code 662, Greenbelt, MD 
20771}

	\altaffiltext{1}{e-mail: barron@astro.spa.umn.edu}
	\altaffiltext{2}{e-mail: larry@astro.spa.umn.edu}
	\altaffiltext{3}{e-mail: gotthelf@gsfc.nasa.gov; Universities
                     Space Research Association} 
	\altaffiltext{5}{e-mail: jonathan@cassiopeia.gsfc.nasa.gov;
	Current address: The North Carolina School of Science \& Mathematics
	1219 Broad St., Durham, NC  27715-2418}

\begin{abstract}
We present the first X-ray expansion measurement of the SNR Cas~A. We 
compared new ROSAT to archived Einstein HRI images with a separation 
of 17 years.  The remnant is seen to be expanding, on average, at a 
rate of $0.20\pm 0.01$ \% yr$^{-1}$, which is twice as fast as the 
expansion of the bright radio ring and two-thirds as fast as the 
ensemble of optical Fast Moving Knots.  This argues that we are 
observing different hydrodynamical structures in each band, although 
all of these components are located on the same patchy ring.  In 
addition, significant variations in expansion rate as a function of 
azimuth around the bright X-ray ring are found.  These findings are 
discussed in the context of two classes of models, involving either an 
ambient medium with a uniform or monotonically decreasing density away 
from the center, or a pre-existing circumstellar shell.
\end{abstract}


\section{Introduction}
Cassiopeia A (Cas~A) is the youngest known Galactic supernova remnant 
(SNR) and is believed to have been caused by a type II supernova in 
$1658\pm3$ (van den Bergh \& Kamper 1983).  This shell-type SNR with a 
$\sim$5\arcmin \ diameter is estimated to be $3.4^{+0.3}_{-0.1}$ kpc 
away (Reed et 
al.  1995).  Young, rapidly evolving SNRs such as Cas~A provide
important laboratories in which to study high energy processes in the
ISM. They are rapidly evolving and therefore allow us to observe
astrophysical hydrodynamics and its coupling to radiation mechanisms
in real time. 

Much analytical and numerical work has gone into studying remnant 
dynamics (e.g., Sedov 1969, Chevalier 1982, Dohm-Palmer \& Jones 1996, 
Borkowski et al.  1996 (B96)).  To test such dynamical models, one 
must identify observed features with the theoretical hydrodynamical 
structures such as shocks or contact discontinuities.  In some 
remnants, e.g. Tycho's SNR (Renoso, et al.  1997), such a connection 
between observed radio features and theoretical hydrodynamics seems 
clear, with the sharp-edged ring being identified with the outer shock 
(Tan \& Gull 1985).  However, the correspondence between Cas~A's 
morphology and dynamics is ambiguous.  In the radio, we have proposed 
(Koralesky \& Rudnick 1998 (KR98)) that the bright ring represents the 
reverse shock or contact discontinuity.  In the X-ray, the bright ring 
has been identified either with the reverse shock (Fabian et al.  
1980) or the interaction region between a circumstellar shell and the 
SNR ejecta (B96).  Therefore, detailed dynamical information in all 
wavebands is needed to discriminate among possible models.  Below, we 
give a brief overview of what is currently known about the structures 
and motions of the optical, radio and X-ray emitting components of 
Cas~A.

The kinematics of small and large scale radio features of Cas~A are 
well known (e.g., Tuffs 1986, Anderson \& Rudnick 1995 (AR95), KR98).  
The kinematics of the small-scale radio knots are significantly 
decelerated relative to the optical emission and inconsistent with a 
homologous expansion.  KR98 examined the motions of the larger-scale 
filamentary bright radio ring and found its expansion to vary by over 
a factor of 2 as a function of azimuth.

The optical components of Cas~A also have very disparate expansion 
rates.  The quasi-stationary flocculi (QSFs) have velocities less than 
about 500 km s$^{-1}$.  The fast moving knots (FMKs) have velocities 
ranging from 4000-6000 km s$^{-1}$, with an average velocity of 5300 
km s$^{-1}$ (Reed 1994).  Fesen, Becker, \& Goodrich (1988) have found 
velocities exceeding 10,000 km s$^{-1}$.  Not only do the different 
types of optical knots distinguish themselves by their kinematics, 
they have different abundances as well.  The QSFs are thus thought to 
be shocked material in the shell and the FMKs ejecta from the SN 
explosion.

A velocity asymmetry was first measured in Cas A's X-ray line of sight 
doppler velocities by Markert et al.  (1983).  Later, Holt et al.  
(1994) using the ASCA X-ray observatory found a 2100 km s$^{-1}$ 
velocity asymmetry in Cas A, receding in the NW and approaching in the 
SE. It is likely that this is a result of the patchy, irregular 
illumination of Cas A's shell, since the same asymmetry is seen in the 
radio rotation measures (Kenney \& Dent 1985, Rudnick \& Koralesky 
1998 (RK98)).  Thus, the X-ray velocity gradient may result from the 
far side being seen in the NW and the near side seen in the SE. The 
distribution of FMKs is also very patchy on large scales (Reed et al.  
1995, Lawrence et al.  1995).

Looking at the overall remnant structure, Reed et al.  (1995) found 
that 76\% of the FMKs fall within 5\arcsec \ of a spherical shell 
which is at the same radius as the radio and X-ray rings.  A strong 
correlation between the X-ray and radio morphology has been 
demonstrated by Keohane, Rudnick, \& Anderson (1996, KRA96) and 
Keohane, Gotthelf, \& Petre (1998, KGP98).  KGP98 suggest that this 
correlation is consistent with proportionate partition between the 
local magnetic field and the hot gas, which would imply turbulent 
mixing of the components.

Despite the overlap between the emission in the different wavebands, 
it is clear that there are multiple dynamical components.  Even within 
the X-ray waveband, Cas~A's spectra have been modeled by two thermal components 
which have been traditionally interpreted to be outer- and 
inner-shocked ejecta (Markert et al.  1983, Jansen et al.  1988).  The 
majority of Cas~A's X-ray emission has been attributed to the 
inner-shocked ejecta (Fabian et al., 1980), which has been shown to be 
enriched in S and Ar as are the FMKs (Vink, Kaastra, \& Bleeker 1996).  
More recently, Allen et al.  (1997), have found strong evidence for a 
high-energy tail in Cas~A that dominates the X-ray spectrum above 10 
keV. They argue that this is due to synchrotron radiation of electrons 
accelerated to energies of 40 TeV.

The kinematics of Cas A's X-ray component are a crucial piece of 
information in disentangling the complex behavior of this remnant.  
The X-rays may be linked to the behavior in other wavebands, or may 
probe an entirely different hydrodynamical structure.  For instance, 
in the case of Tycho, Hughes (1996) found that the X-ray dynamics were 
distinct from the behavior in the radio or optical.

\section{Observations \& Analysis}

We observed Cas~A with two new pointings, taken with the high 
resolution imager (HRI) on the ROSAT X-ray satellite, on 23 Dec. 1995 
and 20 Jun. 1996 with a combined yield of 232 ks of acceptable time at an 
effective epoch of 1996.1.  The two images were aligned and registered 
with the radio image as described in KGP98 and shown as contours in Figure 1.  
Note the lower intensity in the SW, which is due to absorption of the 
soft X-rays by a molecular cloud (KRA96).  We compared this with an 
archived Einstein HRI exposure of Cas~A from 1979 (Fabian et al.  
1980, Murray et al.  1979).  The 17.0 year baseline between Einstein 
and ROSAT observations represents $\sim$5\% of the age of the remnant.  
Cas~A is imaged in the center of the field on the optical axis where 
the resolution is optimal and the distortion minimal.  The HRIs on 
Einstein and ROSAT are well suited for comparisons of this type 
because of their similar design and spectral response in their overlapping 
bandpasses.  The measured  flux from Cas~A is dominated by the 
1.86 keV Si line, which provides the bulk of the emission over the 
bandpass of both instruments.  Both maps were normalized by their 
total counts and convolved with a 5\arcsec \ Gaussian to put them on 
the same spatial and flux scale.

The radio expansion measurements were taken from KR98, which presents 
a more complete discussion of the uncertainties and possible biases in 
expansion analysis presented herein.  The radio data were taken at the 
Very Large Array\footnote{The National Radio Astronomy Observatory is 
a facility of the National Science Foundation operated under 
cooperative agreement by Associated Universities, Inc.} from epochs 
1983.78 (observers: R.J. Tuffs, S.F. Gull, R.A. Perley) and 1994.58 
(KR98) at an average frequency of 4.8 GHz.

Since optical and radio studies have shown that the motions are 
dominated by expansion and also vary azimuthally, we decided to focus 
our initial study of X-ray motions in those areas.  We segmented each 
X-ray image into sectors of 10 degrees, and compared spatially 
expanded versions of the 1979.1 image, sector by sector, with the new 
1996.1 image in the same manner as the radio.  An expansion center of 
RA(1950) 23$^{\rm h}$ 21$^{\rm m}$ 11.4$^{\rm s}$ \ Dec (1950) 58\arcdeg 
32\arcmin 28.5\arcsec \ was assumed for consistency with the radio 
analysis.  We used a range of expansion factors from 0.97 to 1.08.  
The ``observed'' expansion factor for each sector was determined as 
the one that produced the minimum rms difference between the 
normalized Einstein and ROSAT sector images.  All analysis was done 
using NRAO's AIPS. We then corrected the observed expansion factors for the 
0.50\% plate scale differences between ROSAT and Einstein (David et 
al.  1995).

The ring varies in width from 5-20\arcsec \ in the radio and from
10-20\arcsec \ in X-rays.  In most sectors, the contributions from the
bright ring dominate the determination of the expansion in both these
wavebands.  

Furthermore, we  measured the 
relative radii of the rings between the X-ray and radio images using 
the above expansion determination method.  For 
the ROSAT image, we used a comparison radio image interpolated to the 
same epoch using maps from KR98.  For the Einstein image, we used the 
1979 radio map from the Cambridge 5 km interferometer (Observer: R. J. 
Tuffs).  We then minimized the rms difference as a function of 
``expansion'' factor between each X-ray map and its respective radio 
counterpart, and thus determined the relative radii of the bright 
rings at each epoch.

\section{Results}

The observed and corrected expansion factors for Cas~A's X-ray 
emission as a function of azimuth are shown in Figure 2, along with 
the radio expansion values from KR98.  Both have been converted onto a 
scale of percent expansion per year (hereinafter ``expansion rate'').  
For this plot, we smoothed the data using a boxcar of 30$^{\circ}$, so 
there are three (five) independent points for each X-ray (radio) 
plotted point.  The most striking result is that the median expansion 
rate in the X-ray, $0.20\pm0.01$ \% yr$^{-1}$, is twice that observed
for the radio ring (KR98).  These  
translate to an expansion parameter, $m=0.73$ and 0.35 respectively, 
here $R\propto t^{m}$, where $R$=radius to the ring and $t$=time.  
For reference, these expansions correspond to $\sim3500$ and 
$\sim1750$ km s$^{-1}$, respectively.  This velocity is larger
than measured by Holt et al. (1994).

We estimated the errors using  ($\frac{1}{\sqrt{2}}$, X-ray) and 
($\frac{1}{2}$, radio) times the rms scatter among the independent 
sector values going into each plotted point in Figure ~1.   While this
gives only an approximate measure of the actual errors, it does
include all contributing factors including the statistical error on
the count rate and structural influences to the fit accuracy.

There are also X-ray expansion rate variations of $\sim$50\% as a 
function of azimuth.  However, any differences between the center used 
and the unknown explosion center would result in a sinusoidal 
variation with azimuth in the observed expansion rate.  Also, registration 
offsets between the Einstein and ROSAT images would mimic this effect.
For example, a 
position offset of 5\arcsec (consistent with a mis-alignment of 
Einstein and ROSAT by one convolved beam), we would expect a sinusoid 
of magnitude 0.3\% yr$^{-1}$.  It is therefore likely that some of the 
large scale azimuthal variations in Figure 2 are due to such errors.  
Large angular scale variations are also seen in the radio data; these 
cannot be due to registation uncertainties and are likely the result from 
different deceleration in the presence of external density gradients 
(AR95).

There are also significant smaller angular scale variations, such as 
the marginal, but sharp dips at azimuths $\sim 170^{\circ}$ and $\sim 
250^{\circ}$ which are seen in both the X-ray and radio measurements.  
These cannot 
be caused by position offset biases due to their smaller angular scale.  
In addition, there appears to be a very sharp drop in X-ray expansion 
rate at $65^{\circ}$.  This is the region commonly known as the ``jet'',
where the bright ring is only marginally detected in either the radio
or X-ray. 

In comparing the relative radii of the X-ray and radio rings, we 
measured the mean X-ray size to be smaller than the radio by $0.16\% \pm
0.06$ in 1996.1 and by  $3.6\% \pm 1.1$ in 1979.  For reference, these
correspond to $\sim0.003$ and $\sim0.06$ pc respectively.  It appears
that the X-ray ring is ``catching up'' to the radio.  

\section{Discussion}

The dramatically different mean X-ray and radio expansion rates argue 
that we are observing two different hydrodynamical structures in these 
bands.  Their emission shows that these structures are also dominated
by different radiative processes.  The  
X-ray brightness traces the regions of hot, dense gas and is sensitive 
to ionization state, while the radio would follow the regions of 
enhanced magnetic field.  These populations are not necessarily 
coupled to the same magneto-hydrodynamic structures in a young SNR.

Several conclusions can be drawn in the context of two existing models 
of Cas~A.  
For the purposes of this discussion, we will assume that we 
are observing a well-defined X-ray component.  Since the X-ray spectra 
show more than one component, future work may show these are also 
distinct kinematically, and whether we are observing a blend.  The first 
model is that of a SNR expanding into a uniform or monotonically 
decreasing density gradient.  The outer shock cannot be the origin of 
either X-ray or radio rings, because the large-scale diffuse plateau 
of radio and X-ray emission exists outside of both bright rings.

The X-ray ring can also not be tracing the reverse shock during the 
early phase of the SNR evolution.  During that time, the reverse shock 
is still expanding inward in the frame of the ejecta.  Therefore the 
ejecta, which the radio is likely tied to, would be moving faster than 
the reverse shock in the external reference frame.  However, the X-ray 
motions could be consistent with the later behavior of the reverse 
shock, once it has reflected from the explosion center and moves 
outward through the ejecta (Dohm-Palmer \& Jones 1996).  During that 
period, the reverse shock will expand more rapidly in the external 
frame than the ejecta, so might reasonably expand more rapidly than 
the radio structures generated by magnetic field amplification in the 
ejecta.  This consistently explains both the higher velocity X-ray 
component and that it could be interior to the radio and catching up 
to it.

Chevalier \& Liang (1989) developed a model where the remnant is
dominated by SN ejecta interacting with the pre-existing circumstellar
shell which was created through wind interactions during the late
stages of evolution of the pre-SN star.  Borkowski et al.  (1996),
propose that 
the outer shock has already been transmitted through the shell and is 
now propagating into the unshocked red supergiant wind, while the 
reverse shock is traveling into the ejecta.  In the context of this
model, the X-ray- and radio-emitting regions trace the shocked diffuse
component of the shell, while the optical knots follow the shocked clumps
which also existed in the shell. 

In its simplest form, the circumstellar shell model does not explain 
the different velocities and positions of the X-ray and radio rings.  
However, it may be tenable in the case of shocks reflected within the 
shell as proposed by B96.  Also, it could be consistent with the 
passage of the reflected inner shock as discussed above, or in the 
case of multiple shocks due to an inhomogeneous medium and viewed in 
projection.

The strong structural correlation between the X-ray and radio observed 
by KGP98 is somewhat puzzling given the kinematic differences 
presented here.  This would seem to imply that we are observing the 
system at almost exactly the right moment when the features are 
overlapping.  However, the structural and kinematic determinations 
actually depend on different X-ray and radio features.  KGP98 find 
their strongest correlation in low-surface-brightness regions while 
our method more effectively tracks the bright ring where they find the
most scatter.

\section{Conclusions \& Future Work}

The kinematics of Cas~A in the X-ray present an intriguing puzzle.  
The average X-ray expansion rate rests in the middle of the average 
radio expansion and the average fast moving optical knot velocity.  
All of these vary significantly as a function of azimuth.  To add to 
the complexity, each of these components exist at the same approximate 
location in the remnant.  Although each observed waveband reveals a 
separate dynamical component, the link between them has yet to be 
determined.  As a next step, we will analyze the two-dimensional 
velocity distribution in the X-rays, to look for non-radial motions 
such as are seen in the radio (AR96).

Higher resolution X-ray instruments, such as AXAF, will be important 
tools to decipher the conditions in Cas~A.  The high spatial and 
spectral resolution of AXAF will allow for more detailed mapping 
of  doppler velocities and spatial variation of abundances.  
By analogy with the 
optical data, the X-ray emission may divide into separate composition 
and kinematic populations.  In addition, follow-up observations will 
allow us to see whether the X-ray and radio shells diverge, as the 
velocities presented suggest they should, or whether a new population 
of X-ray emitting features arises in the interaction region, as is 
probably happening with the optical knots.  This may enable 
differentiation between the two types of models discussed above.  
Careful comparison of the new X-ray and radio measurements 
with the mid-IR, e.g. ISO (Tuffs 1998),  will help trace the
distributions of SN ejecta condensates.

There is also the intriguing  possibility of spatially separating the
two thermal and the relativistic components, all of which could be
blended here.  B96 noted that circumstellar shell models would give a
denser, cooler shell, surrounded by a hotter, less dense shocked wind
gas.  This is the opposite of what one would expect from an expansion
into a uniform medium. 

This work provides an excellent basis for comparison to recent
advancements in multi-dimensional magneto-hydrodynamic  
simulations which will be a crucial component of deciphering Cas~A's 
dynamics.  This, coupled with improved non-equilibrium ionization 
models, will yield more realistic simulations to compare with 
observations.

\bigskip
{\bf Acknowledgments}

SNR research at the University of Minnesota is supported by the NASA 
Graduate Research Program and the National Science Foundation under 
grant AST 96-19438.  We would like to thank T. W. Jones and S. P. 
Reynolds for discussions about SNR dynamics and comments on the paper 
and R. Petre for assistance with the ROSAT observations.

\bigskip
{\bf References}
\bigskip

Allen, G. E. et al. 1997, ApJ, 487, L97

Anderson, M. C., Jones, T. W., Rudnick, L., Tregillis, I. L., \& 
Kang, H.  1994, ApJ, 421, L31 

Anderson, M. C.  \& Rudnick, L.  1995, ApJ, 441, 307 

Borkowski, K. J., Szymkowiak, A. E., Blondin, J. M., \& Sarazin, C. 
L., ApJ, 466, 866

Chevalier 1982, ApJ, 259, L59

Chevalier \& Liang, 1989

David, L. P., Harnden, F. R., Kearns, K. E., \& Zombeck, M. V. 1995,
The ROSAT High Resolution Imager (HRI)

Dohm-Palmer \& Jones 1996, ApJ, 471, 279

Fabian, A. C., Willingale, R., Pye, J. P., Murray, S. S., \& 
Fabbiano, G. 1980, MNRAS, 193, 175

Fesen, R. A., Becker, R. H., \& Goodrich, R. W. 1988, ApJ, 329, L89

Holt, S. S., Gotthelf, E. V., Tsunemi, H., \& Negoro, H. 1994, PASJ, 
46, L151

Hughes, J. 1996, BAAS, 189, 4608

Jansen, F., Smith, A., Bleeker, J. A. M., de Korte, P. A. J., Peacock, 
A., \& White, N. E. 1988, ApJ, 331, 949

Kenney, J. D., \& Dent, W. A. 1985, ApJ, 298, 644

Keohane, J. W., Gotthelf, E. V., \& Petre, R. 1998, ApJ, accepted

Keohane, J. W., Rudnick, L., \& Anderson, M. C. 1996, ApJ, 466, 309

Koralesky \& Rudnick, 1998, in prep.

Lawrence et al. 1995, AJ, 109, 2635

Markert, T. H., Canizares, C. R., Clark, G. W., \& Winkler, P. F. 
1983, ApJ, 268, 134

Murray, S. S., Fabbiano, G., Fabian, A. C., Epstein, A., \& Giacconi, 
R. 1979, ApJ, 234, L69

Reed, J. E., Hester, J. J., Fabian, A. C., \& Winkler, P. F. 1995, 
ApJ, 440, 706

Reed, J. E., Hester, J. J., \& Winkler, P. F. 1995, preprint

Renoso, et al. 1997, ApJ, 491, 816

Rudnick, L. \& Koralesky, B. 1998 preprint

Sedov, L. I. 1969, Similarity and Dimensional Methods in Mechanics 
(New York: Academic)

Tan, S. M., \& Gull, S. F. 1985, MNRAS, 216, 949

Tuffs, R. J. 1986, MNRAS, 219, 13

Tuffs, R.J. 1998, `Observations of Supernova Remnants with ISO', to appear
in `Highlights of the ISO Mission', ed. D. Lemke, in
`Highlights of Astronomy', Proc. XXIIIrd General Assembly of the IAU

van den Bergh, S., \& Kamper, K. 1983, ApJ, 268, 129

Vink, J., Kaastra, J. S., \& Bleeker, J. A. M., 1996, A\&A, L44

%%%%%%%%%%%%%%
% Figures
%%%%%%%%%%%%%%

\begin{figure}
	\vskip 4.in
%	\vbox to7.0in{\rule{0pt}{2.6in}}
	\includegraphics{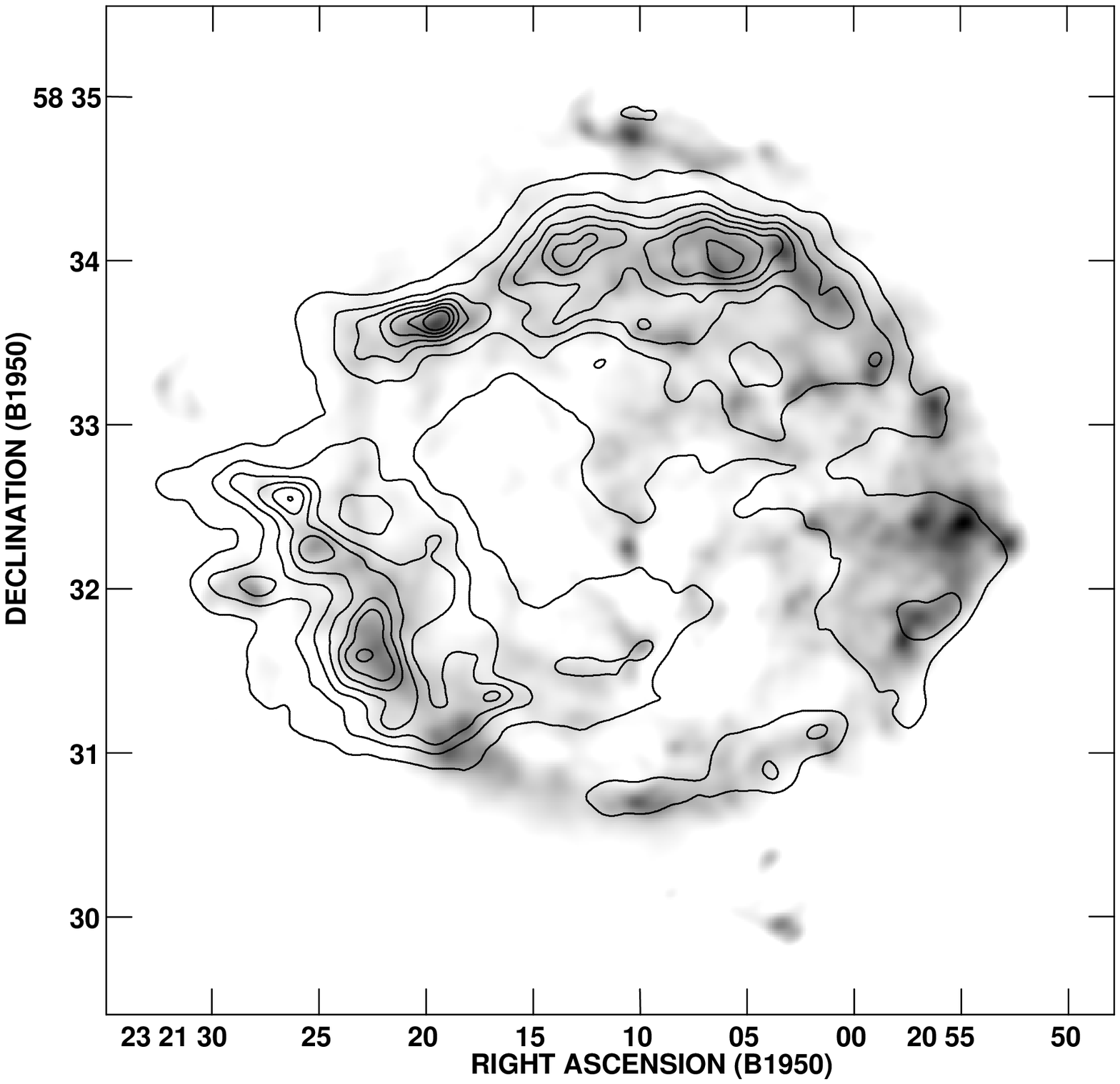}	
\caption{Radio continuum image of Cas~A at 4.8 GHz with X-ray 
emission, as seen by ROSAT, overlayed in contours.  Note the large 
scale similarity between the two observations and that they are 
overlapping on the bright ring (radius $\sim100$\arcsec).  Both are 
convolved with a 5\arcsec \ Gaussian.  Greyscale range is 0.2 to 1.3 
Jy/Beam, and contour levels are 450, 900, 1350, 1800, 2250, 2700, and 
3150 counts.}
\label{CasX-R}
\end{figure}

\begin{figure}
	\vskip 3.5in
%	\vbox to7.0in{\rule{0pt}{2.6in}}
\includegraphics{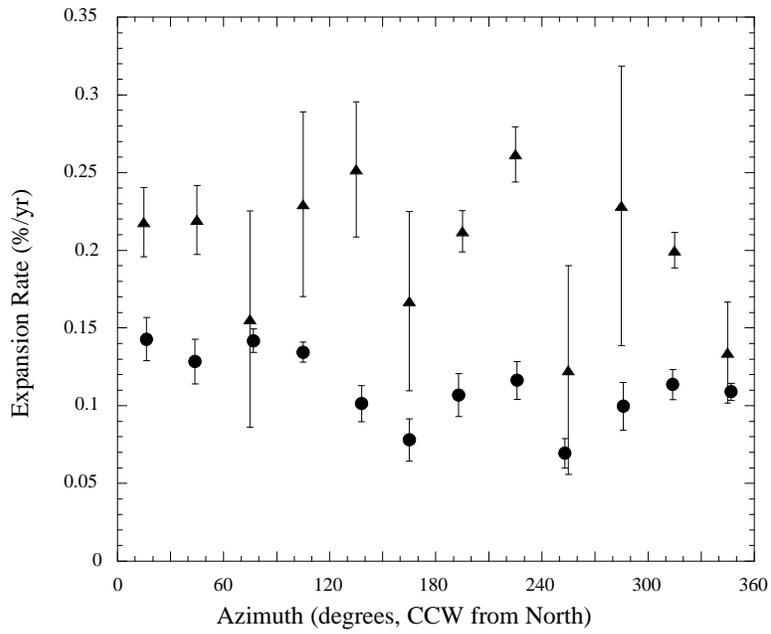} 
\caption{Expansion rate vs. azimuth for both the radio (circles) and X-ray 
(triangles) components of Cas~A. Average expansion rate in the X-ray (0.20\% 
yr$^{-1}$) is {\it twice} that found the radio.  The 
data have been smoothed by 30$^{\circ}$ in azimuth.  Errors represent the
rms scatter in each 30$^{\circ}$ bin (see text)}
\label{EFvAz}
\end{figure}

\end{document}